# *CHANDRA* MULTIWAVELENGTH PROJECT: NORMAL GALAXIES AT INTERMEDIATE REDSHIFT

D.-W. Kim[1], W. A. Barkhouse[2], E. R. Colmenero[3], P. J. Green[1], M. Kim[1],
A. Mossman[1], E. Schlegel[1], J. D. Silverman[4], T. Aldcroft[1], C. Anderson[1],
Z. Ivezic[5], V. Kashyap[1], H. Tananbaum[1], B. J. Wilkes[1]

1. Smithsonian Astrophysical Observatory, Cambridge MA
2. Department of Astronomy, University of Illinois at Urbana-Champaign
3. South African Astronomical Observatory, South Africa
4. Max-Planck-Institut fur extraterrestrische Physik, Garching, Germany
5. Department of Astronomy, University of Washington, Seattle WA

## ABSTRACT

We have investigated 136 Chandra extragalactic sources without broad optical emission lines, including 93 galaxies with narrow emission lines (NELG) and 43 with only absorption lines (ALG). Based on $f_X/f_O$, $L_X$, X-ray spectral hardness and optical emission line diagnostics, we have conservatively classified 36 normal galaxies (20 spirals and 16 ellipticals) and 71 AGNs. Their redshift ranges from 0.01 to 1.2, with normal galaxies in the range $z$=0.01-0.3. Our sample galaxies appear to share characteristics with local galaxies in terms of X-ray luminosities and spectral properties, as expected from the X-ray binary populations and the hot interstellar matter (ISM). In conjunction with normal galaxies found in other surveys, we found no statistically significant evolution in $L_X/L_B$, within the limited $z$ range ($\lesssim 0.1$). We have built a log(N)-log(S) relationship of normal galaxies in the flux range, $f_X$ (0.5-8 keV) = $10^{-15} - 10^{-13}$ erg s$^{-1}$ cm$^{-2}$, after correcting for completeness based on a series of simulations. The best-fit slope is -1.5 for both S (0.5-2 keV) and B (0.5-8 keV) energy bands, which is considerably steeper than that of the AGN-dominated cosmic background sources, but slightly flatter than the previous estimate, indicating normal galaxies will not exceed the AGN population until $f_X$ (0.5-2.0 keV) ~ 2 x $10^{-18}$ erg s$^{-1}$ cm$^{-2}$ (a factor of ~5 lower than the previous estimate). We have also built an X-ray luminosity function of normal galaxies in the luminosity range of $L_X$ = 5 x $10^{39} - 10^{42}$ erg s$^{-1}$, which is consistent with other survey results.

A group of NELGs (most of them with $f_X/f_O$>0.1) appear to be heavily obscured in X-rays, i.e., a typical type 2 AGN. After correcting for intrinsic absorption, their X-ray luminosities could be $L_X > 10^{44}$ erg s$^{-1}$, making them type 2 quasar candidates. While most X-ray luminous ALGs (XBONG – X-ray bright, optically normal galaxy candidates) do not appear to be significantly absorbed, we found two heavily obscured objects, which could be as luminous as an unobscured broad-line quasar. Among 43 ALGs, we found two E+A galaxy candidates with strong Balmer absorption lines, but no [OII] line. The X-ray spectra of both galaxies are soft and one of them has a nearby close companion galaxy, supporting the merger/interaction scenario rather than the dusty starburst hypothesis.

*Subject headings*: surveys – X-rays: galaxies – X-rays: general

## 1. INTRODUCTION

To understand the formation and evolution of galaxies, it is important to study galaxies at a wide range of redshifts starting from the time when they form and rapidly evolve. The X-ray observations of normal galaxies at high $z$ provide valuable information, which is directly related to star formation episodes: hot bubbles/winds and high-mass X-ray binaries (HMXB) reflecting on-going star formation or young stellar populations, and low-mass X-ray binaries (LMXB) reflecting old stellar populations. In particular, Ghosh and White (2001) predict that the X-ray luminosity at higher $z$ could be 10-100 times higher than that in the local galaxies with HMXBs peaking at $z$=1-2 while LMXBs at $z$=0.5-1.0 due to the delayed turn-on.

Recent *Chandra* and *XMM-Newton* X-ray data have dramatically increased our understating of both distant high-z galaxies and local galaxies. Investigating the Chandra Deep Fields (CDF), Hornschemeier et al. (2003) identified a significant number of normal galaxies out to cosmologically significant distances and showed that normal galaxies outnumber AGNs at faint fluxes (see also Bauer et al. 2004). Norman et al. (2004) have determined an X-ray-derived star formation rate and suggested mild evolution with star formation rate ~ $(1+z)^{2.7}$. On the other hand, nearby galaxies are investigated in detail with high resolution X-ray images, primarily because point sources, either LMXB or HMXB, are individually detected both in elliptical galaxies (e.g., Sarazin et al. 2001) and late type star forming galaxies (e.g., Colbert et al. 2003). X-ray luminosity functions of X-ray binaries are built (e.g., Grimm et al. 2003; Kim & Fabbiano 2004) and their implications to star formation histories and to compact binary formation models are being untangled (e.g., Gilfanov et al. 2004; Belczynski et al. 2004).

The *Chandra* Multiwavelength Project (ChaMP; Kim et al. 2004; Green et al. 2004), with the advantage of a wide area coverage, allows us to investigate a well-defined galaxy sample at redshifts intermediate between distant galaxies found in CDF and local galaxies. In this work, we combine our ChaMP galaxy data with other data from the CDF and XMM surveys to obtain a complete picture covering a wide range of redshifts and apply our knowledge (e.g., $L_X/L_O$) obtained from local galaxies to the higher $z$ sample.

This paper is organized as follows. In section 2, we describe our sample selection, classification schemes and sample characteristics. In section 3, we discuss normal galaxies in terms of their $L_X/L_B$ evolution, Log (N)-Log (S) and X-ray luminosity function. In section 4, we present obscured type 2 AGNs and quasars. In section 5, we discuss an unusual population of X-ray bright optically normal galaxies (XBONG). In section 6, we present two E+A galaxy candidates and discuss the origin of this mysterious phenomenon. Finally, we summarize our conclusions in section 7.

Throughout this paper, we use $H_o$=70 km s$^{-1}$ Mpc$^{-1}$, $\Omega_m = 0.3$ and $\Omega_\Lambda$=0.7.

## 2. ChaMP GALAXY SAMPLE

### 2.1. Sample Selection

ChaMP fields (130 from Observing cycles AO1 and AO2) were selected by various criteria to optimize for extra-galactic sources (Kim et al. 2004a). Among those X-ray sources detected in 47 fields with follow-up ChaMP optical imaging and spectroscopic observations, we have selected galaxies based on optical spectroscopic identification (Green et al. 2004; Silverman et al. 2005). Our sample includes narrow emission line galaxies (NELG) and absorption line galaxies (ALG), but excludes broad line AGN/QSO (with a line width < 1000 km s$^{-1}$) and galactic stars. To ensure our identification, we restrict our sample to 105 sources, which have both a clear optical counterpart (i.e., highest match confidence in the ChaMP database) and a reliable optical spectroscopic identification (i.e., highest type confidence in the ChaMP database). We refer to Kim et al. (2004a) and Kim, M. et al. (2005, in preparation) for the detailed description of the X-ray data analysis and to Green et al. (2004) and Silverman et al. (2005) for the optical data analysis and spectroscopic identification. For additional details, we also refer to the ChaMP public web page (hea-www.cfa.harvard.edu/CHAMP).

We have also cross-correlated ChaMP X-ray sources with SDSS DR3 optical sources (www.sdss.org/dr3) and found that 31 additional objects satisfy the same selection criteria described above. A full report of the SDSS cross-correlation will be presented in a forthcoming paper. Assuming a small error in SDSS optical source positions, for the match radius we have only applied an X-ray centroiding error, which is formulated as a function of off-axis distance and source counts in Kim et al. (2004a) and Kim, M. et al. (2005). After matching using an automatic script, we visually inspected each source for confirmation by the same technique as with ChaMP optical imaging. We note that ChaMP uses the same optical photometric system as the SDSS (Green et al. 2004).

The final sample consists of 136 galaxies with roughly a 2:1 ratio between NELG and ALG. For six bright galaxies (all ellipticals) for which our ChaMP optical CCD photometry was saturated, we have collected corresponding, but less accurate, optical magnitudes from NED (nedwww.ipac.caltech.edu), and marked them as such in the following figures and Table 1. We have excluded targets (9 sources) of selected Chandra observations to avoid any systematic effect.

### 2.2. Classification

Although we exclude broad line (type 1) AGNs and QSOs, galaxies in our sample still contain a significant contribution from the AGN (type 2) activity to their X-ray and/or optical emission. Therefore, to further separate AGNs from normal galaxies, we utilize several diagnostics: the X-ray luminosity, X-ray-to-optical flux ratio, X-ray spectral hardness and optical line ratios.

An X-ray luminosity of $L_X = 10^{42}$ erg s$^{-1}$ is often used to distinguish the AGN and star forming activities, because the most luminous star forming galaxy has $L_X \sim 10^{42}$ erg s$^{-1}$ (e.g., Moran et al. 1999). Similarly, most AGNs have X-ray-to-optical flux ratios between $f_X/f_O = 0.1$ and 10, as first recognized in the *Einstein* survey by Maccacaro et al. (1988), whereas normal galaxies have $f_X/f_O < 0.01$ (e.g., Kim, Fabbiano & Trinchieri 1991; Shapley, Fabbiano & Eskridge 2001). To be consistent with the original definition by Maccacaro et al. (1988), we define $\log f_X/f_O$ as:

$$\log f_X/f_O = \log f_X(0.5\text{-}8 \text{ keV}) + 5.31 + r/2.5 \text{ or}$$
$$\log f_X/f_O = \log f_X(0.5\text{-}2 \text{ keV}) + 5.71 + r/2.5.$$

We have converted X-ray fluxes (corrected for Galactic absorption) in different energy bands, assuming $\Gamma_{ph}=1.7$ and the Galactic line-of-sight $N_H$ (Dickey and Lockman 1990). We assume 0.2 mag difference between Johnson V and SDSS r magnitudes. Our definition ensures that $f_X/f_O$ remains the same, regardless of the X-ray energy band and optical band used in this formula, except for the small effect caused by the variation of X-ray spectral shapes and optical colors. We note that heavily absorbed X-ray sources (e.g., type 2 AGN) are often very weak in the soft X-ray band with very low X-ray-to-optical flux ratios, if determined in the soft band (0.5-2.0 keV). We therefore use X-ray-to-optical flux ratios determined in the broad band (0.5-8 keV) throughout this paper, unless explicitly mentioned (see also section 4 and 5).

TABLE 1
Number of X-ray sources with different $f_X/f_O$ or $L_X$

| $f_X/f_O$ (or $L_X$) | # of sources determined in S-band | B-band | note |
|---|---|---|---|
| by $L_X$ | | | |
| $\log L_X < 42$ | 81 (51+30) | 51 (29+22) | mostly galaxies |
| $\log L_X > 42$ | 55 (42+13) | 85 (64+21) | mostly type 2 AGN and XBONG |
| total | 136 | 136 | |
| by $f_X/f_O$ | | | |
| undetermined | 3 ( 3+ 0) | 0 | heavily absorbed type 2 AGN/QSO |
| <0.01 | 38 (20+18) | 36 (20+16) | galaxies |
| 0.01-0.1 | 40 (27+13) | 29 (16+13) | mixed |
| 0.1< | 55 (43+12) | 71 (57+14) | AGN2 + XBONG |
| total | 136 | 136 | |

In this study, we primarily use $f_X/f_O$ to distinguish normal galaxies and AGNs: normal galaxy if $f_X/f_O < 0.01$ and AGN if $f_X/f_O > 0.1$. In Figure 1, we plot $r$ against $f_X$ and $f_X/f_O$ against $L_X$. We mark $f_X/f_O = 0.1$ by a dashed line and $f_X/f_O = 0.01$ by a dotted line. Also marked, by a vertical line in Figure 1b, is $L_X=10^{42}$ erg s$^{-1}$. The normal galaxies with $f_X/f_O < 0.01$, AGNs with $f_X/f_O > 0.1$ and unclassified objects with intermediate $f_X/f_O$ are marked by filled circles, filled squares and open circles, respectively. The red and blue colors indicate ALG and NELG, respectively. The six elliptical galaxies with less accurate optical photometry are marked by open circles with a dot in the center. Also plotted are normal galaxies identified by similar selection criteria ($f_X/f_O < 0.01$, but with a slightly different definition of $f_X/f_O$) from the Chandra Deep Field-North (CDF-N; Hornschemeier et al. 2003) and the XMM-Newton Needle in the Haystack Survey (NHS; Georgantopoulos et al. 2005). They are marked by cyan and green triangles, respectively. X-ray fluxes, luminosities, and $f_X/f_O$ of galaxies from these two samples are converted to be consistent with our sample, as described above.

We summarize the number of sources in different bins of $f_X/f_O$ and $L_X$ in Table 1. We have 36 normal galaxies (20 star forming spiral galaxies among NELGs and 16 elliptical galaxies among ALGs; listed in Table 2); 71 AGNs including 57 NELGs (likely type 2 AGNs) and 14 ALGs (these objects are called X-ray bright optically normal galaxies, or XBONG). We note that no galaxy with $f_X/f_O < 0.01$ is more luminous than $L_X = 10^{42}$ erg sec$^{-1}$ and no AGN with $f_X/f_O > 0.1$ is less luminous than $L_X = 10^{42}$ erg sec$^{-1}$ (see Figure 1b), hence our classification also satisfies the distinction by the X-ray luminosity. However, by this conservative scheme, we could not classify 29 sources (or ~20%) with an intermediate $f_X/f_O$ (16 NELGs and 13 ALGs). We will discuss these unclassified objects further in section 2.3.

TABLE 2
Normal Galaxy Sample

| name | type | z | r (mag) | log$f_X$ (erg s$^{-1}$cm$^{-2}$) | log$f_X/f_O$ | log$L_X$ (erg s$^{-1}$) | HR | 1σ limit (lower upper) |
|---|---|---|---|---|---|---|---|---|
| CXOMP J012358.2-350654 | ALG  | 0.019 | 12.50 | -13.70 | -3.39 | 40.22 | -0.23 | (-0.56  0.15) |
| CXOMP J033757.9-050001 | ALG  | 0.036 | 13.40 | -13.57 | -2.90 | 40.88 |  0.38 | ( 0.28  0.48) |
| CXOMP J033851.8-353538 | ALG  | 0.006 |  9.10 | -11.38 | -2.43 | 41.38 | -0.94 | (-0.94 -0.93) |
| CXOMP J033943.1-352159 | ALG  | 0.062 | 16.64 | -14.23 | -2.27 | 40.73 | -0.25 | (-0.54  0.07) |
| CXOMP J051942.2-454953 | NELG | 0.205 | 17.63 | -14.40 | -2.04 | 41.68 | -1.00 | (-1.00 -0.95) |
| CXOMP J082755.1+292659 | NELG | 0.030 | 15.87 | -13.95 | -2.30 | 40.33 | -0.35 | (-0.77  0.03) |
| CXOMP J083228.0+523620 | NELG | 0.017 | 13.88 | -13.24 | -2.37 | 40.52 | -0.81 | (-0.96 -0.62) |
| CXOMP J084527.4+342508 | NELG | 0.026 | 14.22 | -14.00 | -3.00 | 40.15 | -1.00 | (-1.00 -0.46) |
| CXOMP J102543.9+471935 | ALG  | 0.062 | 14.53 | -13.86 | -2.74 | 41.11 | -1.00 | (-1.00 -0.74) |
| CXOMP J105648.9-033725 | ALG  | 0.182 | 17.17 | -14.32 | -2.14 | 41.65 | -0.74 | (-0.85 -0.60) |
| CXOMP J111809.9+074654 | NELG | 0.042 | 15.33 | -14.28 | -2.84 | 40.34 | -0.78 | (-0.97 -0.54) |
| CXOMP J113206.7+045338 | ALG  | 0.150 | 16.73 | -14.16 | -2.16 | 41.62 | -1.00 | (-1.00 -0.52) |
| CXOMP J114058.4+660626 | NELG | 0.236 | 18.61 | -14.93 | -2.18 | 41.29 | -0.43 | (-0.78 -0.08) |
| CXOMP J114134.0+661351 | NELG | 0.145 | 18.26 | -14.70 | -2.08 | 41.06 | -0.54 | (-0.80 -0.22) |
| CXOMP J122131.4+491036 | NELG | 0.185 | 16.87 | -14.51 | -2.45 | 41.47 | -0.82 | (-0.95 -0.57) |
| CXOMP J122139.7+491955 | NELG | 0.125 | 16.83 | -14.52 | -2.48 | 41.08 |  0.20 | (-0.07  0.46) |
| CXOMP J122814.5+442710 | NELG | 0.023 | 13.93 | -13.52 | -2.63 | 40.56 | -0.19 | (-0.51  0.15) |
| CXOMP J131206.4+424127 | ALG  | 0.180 | 17.32 | -14.89 | -2.65 | 41.06 |  0.41 | ( 0.06  0.73) |
| CXOMP J134407.3-002832 | NELG | 0.102 | 16.24 | -14.54 | -2.74 | 40.88 | -1.00 | (-1.00 -0.15) |
| CXOMP J134421.6+555122 | ALG  | 0.037 | 14.48 | -14.69 | -3.59 | 39.81 | -0.60 | (-0.94 -0.20) |
| CXOMP J134422.4+555703 | NELG | 0.038 | 15.50 | -14.80 | -3.29 | 39.73 | -1.00 | (-1.00 -0.90) |
| CXOMP J134428.3+000146 | ALG  | 0.135 | 16.58 | -14.26 | -2.31 | 41.43 | -1.00 | (-1.00 -0.88) |
| CXOMP J134510.4+560143 | NELG | 0.144 | 16.31 | -14.32 | -2.48 | 41.42 |  0.54 | ( 0.24  0.81) |
| CXOMP J141126.9+521551 | NELG | 0.234 | 18.61 | -15.08 | -2.32 | 41.14 | -1.00 | (-1.00 -0.56) |
| CXOMP J141715.7+445545 | NELG | 0.113 | 16.78 | -14.15 | -2.12 | 41.37 | -0.50 | (-0.88 -0.08) |
| CXOMP J151004.0+074037 | NELG | 0.046 | 15.88 | -13.76 | -2.10 | 40.92 |  0.69 | ( 0.29  0.97) |
| CXOMP J151427.1+363803 | ALG  | 0.162 | 17.09 | -14.43 | -2.28 | 41.42 | -0.68 | (-0.94 -0.36) |
| CXOMP J153447.2+232913 | ALG  | 0.090 | 17.64 | -14.55 | -2.18 | 40.75 | -0.63 | (-0.95 -0.22) |
| CXOMP J171724.8+670911 | NELG | 0.138 | 17.28 | -14.62 | -2.40 | 41.08 |  0.04 | (-0.31  0.38) |

```
CXOMP J221306.5-220724   NELG   0.018   15.31   -14.23   -2.79   39.59    0.07  (-0.28  0.40)

CXOMP J221326.2-220547   ALG    0.018   12.90   -13.42   -2.95   40.45   -0.74  (-0.87 -0.60)
CXOMP J221722.7+002107   NELG   0.095   16.70   -14.48   -2.49   40.88   -0.33  (-0.61 -0.06)
CXOMP J230241.0+083856   ALG    0.041   14.70   -14.56   -3.37   40.01   -0.18  (-0.46  0.08)
CXOMP J230241.2+084117   ALG    0.124   17.46   -14.61   -2.32   40.99   -0.48  (-0.69 -0.23)
CXOMP J230252.1+084134   ALG    0.042   13.20   -13.53   -2.94   41.06   -0.80  (-0.89 -0.70)
CXOMP J230308.0+084233   NELG   0.044   17.15   -14.36   -2.19   40.27    0.12  (-0.07  0.31)
```
---

While AGNs could be heavily obscured in X-rays (e.g., X-ray type 2 AGN), normal galaxies are not expected to be significantly absorbed (e.g., Kim, et al. 1992). Therefore, we further check the X-ray spectra to confirm whether X-ray absorbed (hence X-ray faint) AGNs are mixed in the galaxy sample with $f_X/f_O < 0.01$. We use the X-ray hardness ratio, defined as

$$HR = (H-S) / (H+S),$$

where S and H are net counts in 0.5-2.0 keV and 2.0-8.0 keV and X-ray colors as defined in Kim et al. (2004),

$$C21 = \log C_1/C_2 \text{ and } C32 = \log C_2/C_3,$$

where $C_1$, $C_2$ and $C_3$ are net counts in 0.3-0.9 keV, 0.9-2.5 keV, and 2.5-8.0 keV, respectively. By definition, as the X-ray spectra become harder, HR increases and X-ray colors decrease.

TABLE 3
X-ray hardness and colors

| | $f_x/f_o$<0.01 | | 0.01-0.1 | | >0.1 | |
|---|---|---|---|---|---|---|
| | NELG | ALG | NELG | ALG | NELG | ALG |
| | 20 | 16 | 16 | 13 | 57 | 14 |
| HR | -0.40 (0.52) | -0.53 (0.44) | -0.11 (0.48) | -0.31 (0.48) | -0.09 (0.64) | -0.15 (0.47) |
| C21 | -0.38 (0.31) | -0.04 (0.38) | -0.45 (0.34) | -0.55 (0.44) | -0.55 (0.42) | -0.65 (0.28) |
| C32 | 0.35 (0.48) | 0.36 (0.45) | 0.12 (0.49) | 0.34 (0.37) | 0.09 (0.60) | 0.19 (0.38) |

For faint sources with small numbers of counts, there are many cases where the source may be undetected in one of the energy bands, and in almost all cases the assumption of Gaussian distribution of the errors do not apply. HR and colors then often result in unrealistic values and unreliable errors. We have therefore applied a more sophisticated, Bayesian method (Park et al. 2005) that uses the correct Poisson likelihood values and accurately represents the estimates and the confidence ranges at all counts levels. The algorithm also allows a proper accounting of the relative effective areas across an instrument, which is useful in our case because of the different response between the FI (front-illuminated) and BI (back-illuminated) chips. We convert the HR and color estimates of those sources detected in BI chips to the FI system. Taking into account the ACIS QE degradation which could change the soft band counts by 20% or HR by 0.1 (Kim et al. 2004), we also convert the counts to those that would be obtained at the mid-point (June 2000) within the observation period of our sample (AO1 and 2). Figure 2 plots HR against $f_X/f_O$ with/without errors (1σ) for visibility. The symbols are the same as in Figure 1. The mean and standard deviations of each group are listed in Table 3. While AGNs with $f_X/f_O > 0.1$ are spread out in the whole HR range between HR= –1 and HR = 1, most galaxies with $f_X/f_O < 0.01$ are concentrated toward lower HRs (HR ≲ 0.2). In other words, most heavily absorbed objects (HR > 0.6) are AGNs with $f_X/f_O > 0.1$ (see section 4, 5 for the obscured AGNs). The highest HR among those with $f_X/f_O < 0.01$ is CXOMP J151004.0+074037. It is likely an absorbed type 2 AGN (see below). The remaining objects with $f_X/f_O < 0.01$ are consistent with being galaxies within the given error.

For NELGs, we have measured the emission line strengths and applied emission line diagnostics (e.g., Baldwin, Phillips & Terlevich 1981) to further check whether each NELG has an AGN signature. Emission line parameters (flux, equivalent width, FWHM and corresponding errors) were measured using IRAF **splot** for Hα λ6563, [NII] λ6583, [NII] λ6548, [SII] λ6730, [SII] λ6716, OI λ6300, [OIII] λ5007, [OIII] λ4959, Hβ λ4861, and [OII] λ3727. The continuum was estimated by eye at fixed wavelengths. Blended lines were fit with multiple Gaussian functions. We refit each line 5 times with different yet reasonable continuum placements, and derived the average measurement. In Figure 3, we plot various emission line ratios with the curves separating AGNs and galaxies taken from Kewley et al. (2001). Galaxies with $f_X/f_O < 0.01$, AGNs with $f_X/f_O > 0.1$ and intermediate objects are marked by filled circles, filled squares and open circles, respectively (using the same symbols as in Figure 1). It is clear that AGNs are above the curve while galaxies are below the curve, indicating that the classification by the X-ray-to-optical flux ratio is indeed consistent with optical line ratios. One apparent exception (marked by a large circle in the plot) is a galaxy (CXOMP J151004.0+074037), which has $f_X/f_O$ close to the boundary (log $f_X/f_O$ = -2.097) and the highest hardness ratio (HR=0.7) among the galaxy group (as noted in the above), indicating that it is a type 2 AGN with a considerable amount of obscuration. If corrected for absorption, this object would have log $f_X/f_O$ ~ -1. Objects with intermediate $f_X/f_O$ (open circles) are found on both sides of the curve, indicating that they consist of mixed types of galaxies and AGNs. Since most high-z objects do not have Hα and [NII] line measurements, we have also checked those objects only with [OIII] λ5007 and Hβ and confirmed that all AGNs with $f_X/f_O > 0.1$ are consistent within the measurement error with having EW (5007 Å) > 3 EW (Hβ), which is often used to identify the high excitation state induced by the AGN activity (e.g., Szokoly et al. 2004; Steffen et al. 2004).

### 2.3. Unclassified Objects with Intermediate $f_X/f_O$

Our classification scheme is optimized for the least contamination in both normal galaxy and AGN samples. However, for the same reason, we could not classify 29 objects (or ~20%) which have an intermediate $f_X/f_O$ (16 NELGs and 13 ALGs). This group will include galaxies with weak, obscured, or non-existent contributions from an AGN as seen in Figure 3. If they are galaxies with an insignificant AGN component, NELGs could be distant starburst galaxies (e.g., ULIRGs) with a high star formation rate (~100-1000 $M_\odot$ yr$^{-1}$), whereas ALGs could be giant elliptical galaxies with a large amount of the hot ISM (e.g., NGC 507; Kim and Fabbiano 1995). For example, X-ray bright local elliptical galaxies such as NGC 1399 and NGC 4472 have $f_X/f_O \sim$ 0.01-0.1 and $L_X \sim 10^{42}$ erg sec$^{-1}$. Although rare, ALG could be a central elliptical galaxy in a cluster of galaxies with a large amount of ICM. We found one serendipitous cluster (see Barkhouse et al. 2006).

The X-ray spectral properties of these intermediate systems appear to be intermediate between galaxies and AGNs. As summarized in Table 3, going from galaxies to AGNs, the mean hardness ratio increases while the mean X-ray colors decrease with these intermediate objects in the middle, indicating these objects consist of mixed types.

Among those NELGs with emission lines strong enough to apply emission line diagnostics (9 out of 16), we get mixed results. Some objects are below the boundary curve, while others are above, again indicating that they are mixed types. Unfortunately, different line diagnostics produce different identifications. For example, 7 out of 9 are below the curve in the [OIII]/H$\beta$ – [NII]/H$\alpha$ plot (Figure 3a), while 2 of 7 are below the curve in the [OIII]/H$\beta$ – [SII]/H$\alpha$ plot (Figure 3c). Therefore, in this study we do not attempt to classify these intermediate objects, but we will take into account these unclassified sources in the following discussions. We remark that sources with intermediate $f_X/f_O$ are likely to be mixed and mis-classified in various samples in the literatures.

## 2.4. Sample Characteristics

The distributions of optical/X-ray, flux/luminosity, and redshift of our sample are summarized in Figure 4. To calculate the rest frame *r*-band absolute magnitude ($M_r$) we have applied the K-correction and the evolution correction based on the prescriptions in Poggianti (1997), adopting the E1 model with an e-folding star formation time scale of 1 Gyr for ALGs and the Sc model for NELGs. Although there is an uncertainty in selecting the proper evolution model, this will not significantly affect our result because the net correction is always less than 0.2 mag for the normal galaxy sample and less than 0.3 mag for 95% of the whole sample. The X-ray flux and luminosity are corrected for Galactic absorption. For the rest frame X-ray luminosity, we assume $\Gamma_{ph}$=2 for normal galaxies (Kim, Fabbiano & Trinchieri 1992) and $\Gamma_{ph}$=1.7 for AGNs and apply corresponding K-corrections by multiplying by a factor of $(1+z)^{\Gamma_{ph}-2}$.

In Figure 4, NELG and ALG are distinguished by blue solid and dashed lines, while normal galaxies and AGN are marked by red solid and dashed lines, respectively. The redshift of our sample (in Figure 4e) ranges from $z$=0.01 to $z$=1.2, while the normal galaxies are in $z$=0.01-0.3. Since our sample is primarily limited by the X-ray flux, with a limiting flux of $\sim 10^{-15}$ erg s$^{-1}$ cm$^{-2}$ in the energy range of 0.5-8.0 keV, their X-ray luminosity spans a wide range of $L_X = 10^{39} - 10^{44}$ erg s$^{-1}$, depending on the distance. However, we note that the range of optical luminosity is quite narrow with $M_r$ spanning only ~3 mag, in contrast to the *r* magnitude spanning 10 mag (see below for more discussions).

In our sample, the ratio of NELG to ALG is roughly 2:1. The ratio of AGN to normal galaxies is also roughly 2:1. We note that AGNs are preferentially found in NELG (80% of AGNs are NELGs), while normal galaxies are roughly equally divided into NELG (spirals) and ALG (ellipticals).

There is no significant difference between the NELG and ALG sub-samples, except for NELG having a higher AGN fraction. However, AGN and normal galaxy sub-samples appear to be different such that AGNs are more luminous in X-rays, while normal galaxies are more luminous in the optical. The former trend is a natural consequence of our classification scheme, but the latter is rather surprising since there is no selection effect on optical luminosity, particularly on its upper limit. Note that ~40% of normal galaxies have $M_r$ < -22 mag while ~10% of AGN have $M_r$ < -22 mag. The lower limit in optical luminosity could be a selection effect because it would be harder to measure $z$ for distant, less luminous normal galaxies than AGNs.

It is interesting to note that the AGN sub-sample has a very narrow dynamical range in $M_r$. 85% of AGN are within $M_r$ = -21 ± 1 mag, i.e., 1σ rms scatter being merely a factor of ~2. There is a tight correlation between $L_X$ and $f_X/f_O$ (see Figure 1b), as first reported by Fiore et al. (2003) with a sample of type 2 AGN and galaxies. This linear relationship (see the diagonal line with slope=1) is basically dictated by the narrow range in the optical luminosity, compared to the large range (a factor of ~$10^4$) in the X-ray luminosity. The line corresponds to an object with a fixed $M_r$ = -21 mag. However, we note that normal galaxies do not follow this linear relation (see section 3).

## 3. NORMAL GALAXIES

### 3.1. X-ray Emission mechanisms and expected ranges in $L_X$ and $f_X/f_O$

In elliptical galaxies, the major X-ray sources are the hot ISM and LMXBs from the old stellar populations (e.g., Fabbiano 1989). While the amount of the hot ISM significantly varies from one galaxy to another, the X-ray luminosity of LMXBs is well correlated with the optical luminosity (e.g., Fabbiano, Kim & Trinchieri 1992; Kim & Fabbiano 2004). Applying the completeness correction based on extensive simulations to take into account the undetected population of faint LMXBs, Kim and Fabbiano (2004) established the relationship between the LMXB X-ray luminosity and the total stellar luminosity of the galaxy:

$$L_X(\text{LMXB}) / L_B = 0.9 \pm 0.5 \times 10^{30} \text{ erg s}^{-1} / L_{B\odot},$$

where $L_X$ (LMXB) is the total X-ray luminosity of (both detected and undetected) LMXBs determined in the energy range of 0.3-8.0 keV and $L_B$ is the B-band luminosity in unit of $L_{B\odot}$ (adopting $M_{B\odot}$ = 5.47 mag). If we adopt $B-r$ ~ 1 mag for a typical elliptical galaxy and correct for different energy ranges (assuming $\Gamma_{ph}$ = 2.0; Kim et al. 1992), the above relationship corresponds to log $f_X/f_O$ ~ -3, or log $f_X/f_O$ ~ -3.5 for the 1σ lower limit. In Figure 5, which is the same as Figure 1 but includes only normal galaxies (Table 2), the 1σ lower limit is marked by the red long-dashed line to indicate the minimum $f_X/f_O$ expected from the local galaxies.

In star forming galaxies (spirals, starbursts and mergers), the major X-ray sources are those associated with recent star formation, as the X-ray luminosity is closely correlated with the star formation rate (e.g., Grimm et al. 2003; Gilfanov et al. 2004). They include HMXBs, supernova remnants and the diffuse emission associated with the young star clusters and super-winds. Colbert et al. (2003) found a correlation between $L_X$ (point sources) and $L_B$ among a sample of mixed types of galaxies including ellipticals, spirals, and mergers/irregulars. They found that starburst galaxies (e.g., mergers and irregulars) have higher $L_X$ for their $L_B$ than quiescent spiral galaxies and found a linear relation between min $L_X$ (point) and $L_B$ for those non-starburst galaxies. The linear relation also corresponds to log $f_X/f_O$ ~ -3.5 and plotted in Figure 5 (a blue dashed line). See also Hornschemeier et al. (2005) for comparison between Colbert's relationship and other estimates in star forming galaxies. We therefore take the Kim & Fabbiano (2004) relationship as a min $f_X/f_O$ for elliptical galaxies and Colbert's relationship for later type star forming galaxies.

In Figure 5, also plotted are normal galaxies identified by the similar selection criteria (in section 2 and 3) from the Chandra Deep Field-North (CDF-N; Hornschemeier et al. 2003) and the XMM-Newton Needle in the Haystack Survey (NHS; Georgantopoulos et al. 2005). They are marked by cyan and green triangles, respectively as in Fig 1. ALG and NELG are distinguished by filled and open triangles, respectively (with a dot inside an open triangle indicating a mixed type). The symbols for ChaMP galaxies are the same as in Figure 1. As shown in Figure 4e-f, our sample consists of galaxies in the redshift range $z$=0.01-0.3, intermediate between the CDF ($z$=0.1-1.0) and nearby local galaxies, hence covering an intermediate flux range. Our sample is similar to the NHS ($z$=0.005-0.2), but extends to a slightly higher $z$ and a fainter X-ray flux. Combining them with our sample allows us to cover a wide parameter space and to increase the statistical confidence of our results.

While galaxies from the three different sample are distinct in their X-ray flux due to the different observation depths (or redshift coverage) as seen in Figure 5a, they are well mixed in the $f_X/f_O$ - $L_X$ space (or in the $L_r$ - $L_X$ space) with $L_X = 10^{38} – 10^{42}$ erg sec$^{-1}$ and log $f_X/f_O$ = -3.5 – -2, as seen in Figure 5b (and also Figure 6). This confirms that they are indeed similar types of galaxies (but also implies little variation as a function of $z$ – see the next section for more discussion). Therefore, we use them together in the following analyses. The observed minimum $f_X/f_O$ in all three samples are consistently at log $f_X/f_O$ ~ -3.5 (see Figure 5) and this minimum $f_X/f_O$ is consistent with the lower limit of $L_X$ (point source)–$L_B$ (stellar light) relationship of early type galaxies (Kim and Fabbiano 2004) and of spirals (Colbert et al. 2003). Because none of these three samples limits those objects with lower X-ray-to-optical flux ratios (log $f_X/f_O$ < -3.5), this minimum $f_X/f_O$ appears to be real for normal galaxies and to remain constant as a function of $z$. This indicates that the X-ray emission from X-ray binary populations in our sample galaxies is similar to that expected from nearby normal galaxies (see more in section 3.2).

While the X-ray binary contribution can be understood to be proportional to the stellar optical emission, the diffuse X-ray emission (i.e., the excess X-ray emission above that expected from X-ray binaries) is not linearly proportional to the optical light and varies widely from one galaxy to another (e.g., Fabbiano et al. 1992). Furthermore, there may be excess intrinsic absorption in some active star forming regions (e.g., the Antennae; Zezas et al. 2003). The non-linear relationship between the optical and X-ray luminosity was first established with the *Einstein* database (e.g., Fabbiano et al. 1992). Fabbiano et al. (1992) and later Eskridge et al. (1995) found $L_X \sim L_B^{1.8}$ in a sample of early type galaxies, whereas Shapley et al. (2001) found $L_X \sim L_B^{1.5}$ in a sample of spiral galaxies. The steeper slope (> 1) indicates the presence of a significant amount of the hot ISM in elliptical galaxies (Kim et al. 1992) and in early type spiral galaxies (Fabbiano and Shapley 2002), and possibly a luminosity dependence of intrinsically absorbed X-ray emission regions (affecting the blue more significantly than the X-ray band) in late type spiral galaxies (Fabbiano and Shapley 2002).

In Figure 6, we plot $L_X$ against $L_O$ separately for early type and late type galaxies. Again we use all three samples. We note that a single sample covers only a limited parameter space, particularly for early type galaxies. It is clear that for both types of galaxies, the $L_X$ - $L_O$ relation is steeper than a linear relation, as expected from the local normal galaxies. The best-fit slopes are 1.47 ± 0.13 and 1.12 ± 0.11 for early and late type galaxies, respectively. In fitting the early type galaxies, we did not use those saturated in optical images – marked by an open circle with a dot in the center. While the trend for the $L_X$-$L_O$ relation to be steeper in early type than late type galaxies is the same as local galaxies, in both cases our slopes are flatter than those of local galaxies (1.8 and 1.5; see above). The difference is more significant in the late type galaxies (a significance level of ~3σ) than in the early type galaxies (a significance level of ~2σ).

The apparent discrepancy in late type galaxies may be due to our use of the r band, which is less affected by the optical extinction than the B band. In local spiral galaxies, the optical emission from star forming galaxies may be affected by intrinsic absorption, possibly indicating that $L_X$ is intrinsically proportional to $L_O$ (Fabbiano & Shapley 2002). We tested this possibility by using the g band photometry which is available for our ChaMP data, but not for the CDF and NHS data. We obtained a slightly steeper $L_X$ - $L_g$ relation (the best-fit slope of 1.2 ± 0.2), but still not as steep as that in Fabbiano & Shapley (2002). Another possible reason for the flatter relation is the selection effect imposed in our sample (as well as in CDF and NHS samples) by limiting $f_X/f_O$ < 0.01, so that the most X-ray luminous galaxies (e.g., those with intermediate $f_X/f_O$) would have been excluded. Also the slope may depend on the optical luminosity, because our $M_r$ peak is more luminous by ~1 mag than Fabbiano & Shapley $M_B$ peak. Note that for AGN there is a known decrease in $L_X/L_O$ with $L_O$ (Wilkes et al. 1994; Green et al. 1995).

The steeper slope (> 1) in the $L_X$ - $L_O$ relation is also seen in the $f_X/f_O$ – $L_X$ relation (see Figure 1b and 5b). As discussed in section 2.4, while type 2 AGNs follow a linear $f_X/f_O$ – $L_X$ relation (see also Fiore et al. 2003), normal galaxies do not follow the same linear relation (the solid line in Figure 1b and 5b). Note that the linear relationship does not allow any meaningful relation between the optical and X-ray luminosity, because $L_X \sim$ $L_O^n$ and $f_X/f_O \sim L_X$ simultaneously hold only if $n \gg 1$, if we assume $f_X/f_O = L_X/L_O$, i.e. no K-correction. On the other hand, a linear $L_X$ - $L_O$ relation would make $f_X/f_O$ constant, i.e., independent of $L_X$. With our sample of normal galaxies, we found a self-consistent result that the $f_X/f_O$ – $L_X$ relation is flatter than linear (a slope ~ 0.5), while the $L_X$ - $L_O$ relation steeper than linear (Figure 6).

### 3.2. $L_X/L_B$ Evolution

As the star formation rate peaks at $z=2-3$ (e.g., Madau et al. 1996) and declines afterwards, the fossil record of past star formation imprinted on X-ray binaries could be detected by observing galaxies at high redshift (e.g., White and Ghosh 1999; Ghosh and White 2001). In particular, White and Ghosh (1999) predicted that the X-ray luminosity at higher z could be 10-100 times higher than that in the local galaxies with HMXB peaking at $z=1-2$ while LMXB at $z=0.5-1.0$ due to the delayed turn-on of the LMXB population.

Recent stacking analyses with the Chandra deep field data for spiral galaxy samples produced mixed results. Hornschemeier et al (2002) tested $L_X$ evolution for spiral galaxies at $z=0.4-1.5$ and found their $L_X$ and $L_X/L_B$ are statistically consistent with those of local spiral galaxies (Shapley et al. 2001), although they could not exclude an increase by a factor of 2-3. Lehmer et al. (2005) studied Lyman break galaxies at $z=3-6$ and found that $L_X/L_B$ is elevated at $z\sim3$ over that of local galaxies, but consistent at higher $z$. Applying a similar stacking technique with XMM-Newton data of a sample of spiral galaxies at a mean redshift, $<z>=0.1$, Georgakakis et al. (2003) found no $L_X/L_B$ evolution.

We note that the observational measurement of $L_X/L_B$ and its interpretation in term of the evolution of X-ray binary populations depend critically on both the selection and the non-stellar (i.e., AGN or the hot ISM) contribution to the total X-ray luminosity. An X-ray luminous galaxy, which has a higher $L_X/L_B$ by a factor of 10-100 than the local galaxy as predicted (e.g., Ghosh and White 2001), might be difficult to distinguish from an AGN. Conversely, if normal galaxies were selected by $f_X/f_O$ as in our sample (and the CDF and NHS samples as well), a normal galaxy with a high $L_X/L_B$ (e.g., those unclassified with an intermediate $f_X/f_O$) would have been excluded. Therefore, a simple mean value of $L_X$ or $L_X/L_B$ for a given sample may not reflect the real X-ray properties of normal galaxies.

In Figure 7, we plot the X-ray-to-optical luminosity/flux ratios as a function of $z$. Again we have plotted normal galaxies from the CDF-N and NHS surveys as well. The small difference between the flux and luminosity ratio is caused by the K-correction and the evolution correction (mostly in the optical band), as described in section 2. In an individual sample, the given X-ray flux limit produces an apparent correlation between $L_X/L_B$ and z and makes it impossible to search for any change as a function of $z$. Nonetheless, utilizing all three samples together, we cover a full $L_X$-z space up to $z=0.1$, where selection biases are not significant. Within this limited z range ($z<0.1$ or look back time of ~1 Gyr), the distributions of $L_X$ and $L_X/L_B$ appear to be similar at different redshifts with $L_X = 10^{38} - 10^{42}$ erg s$^{-1}$ and log $L_X/L_B$ = -3.5 – -2, indicating no significant change in the X-ray properties of normal galaxies. More interestingly, the minimum value of $L_X/L_B$ (or $f_X/f_O$) remains constant up to $z=0.1$. Because this minimum value directly reflects the X-ray binary populations as seen in Figure 5a (or section 3.1), the constant minimum of $L_X/L_B$ indicates no luminosity evolution in the X-ray binary population up to $z=0.1$. This is consistent with the previous results by Georgakakis et al. (2003) over a similar redshift range, but our usage of the minimum $L_X/L_B$ rather than the mean value directly relates to the X-ray binary populations. To observationally test the evolution of X-ray binary populations, it is critical to cover the faint $L_X$ (~$10^{38}$ erg s$^{-1}$) at high $z$ ($z=0.5-1.0$), but that requires the next generation X-ray mission.

### 3.3. Normal galaxy log(N)-log(S) relationship

To determine the log(N)–log(S) relationship for normal galaxies, we explore a sub-sample of 24 Chandra observations (covering ~2 deg$^2$) where the ChaMP optical coverage is complete (as of mid 2004) for galaxies which are optically bright with $r < 19$ mag (see Figure 1). This is the same data set used in the ChaMP hard X-ray emitting AGN study by Silverman et al. (2005). To correct the X-ray completeness, we have performed a series of simulations to determine the sky area as a function of $f_X$. This technique was originally developed by Kim and Fabbiano (2003, 2004) to determine an X-ray luminosity function of LMXBs in elliptical galaxies (as briefly described in section 3.1). The technical details applicable to ChaMP sources are presented in Kim, M., et al. (2006). Figure 8 shows our log(N)–log(S), which was built with galaxies with more than 10 net counts individually in the soft and the broad energy band.

Our log(N)-log(S) relations in the S and B bands appear to have the same shape, with a best-fit power-law slope of -1.5 ± 0.15 (i.e., Euclidean). The best-fit power-law distribution is plotted in Figure 8. This is not surprising, because we are dealing with a homogeneous sample of galaxies with a relatively small amount of obscuration. This is in contrast to AGN-dominated cosmic background sources, where obscured and unobscured X-ray sources contribute differently in different energy bands (e.g., Kim et al 2004). We also note that unlike the broken power-law distribution of cosmic background AGNs with a break at $f_X$ (0.5-2.0 keV) = 6 x 10$^{-15}$ erg s$^{-1}$ cm$^{-2}$ (e.g., Kim et al. 2004), the log(N)-log(S) of normal galaxies is well reproduced by a single power-law in a wide range of fX. This trend seems to continue to $f_X$ (0.5-2.0 keV) ~ 10$^{-12}$ erg s$^{-1}$ cm$^{-2}$ (Tajer et al. 2005).

Also plotted in Figure 8 are those previously determined with the CDF data in the S-band (Hornschemeier et al. 2003) and with the NHS data in the B-band (Georgakakis et al. 2004). It appears that our S-band log(N)-log(S) can be connected to that of the CDF, if extrapolated. However, the CDF data point at the faintest $f_X$(0.5-2.0 keV) ~ 3 x 10$^{-17}$ erg s$^{-1}$ cm$^{-2}$ is slightly higher than the extrapolated value and their best-fit slope, -1.74 ( 0.3, although statistically consistent, is slightly steeper than ours. Our B-band log(N)-log(S) is statistically identical with that of the NHS, except our data cover lower fluxes with smaller error bars.

As discussed in Hornschemeier et al. (2003) and Bauer et al. (2004), normal galaxies are the dominant cosmic X-ray background source at fainter flux levels, because the log(N)-log(S) slope is steep, compared to the slope of -0.6 – -0.7 (e.g., Bauer et al. 2004) of cosmic background AGNs at faint fluxes ($f_X < 10^{-14}$ erg s$^{-1}$ cm$^{-2}$, i.e., below the break). The dominance of normal galaxies over the AGN population would occur at $f_X$ (0.5-2.0 keV) ~ $10^{-17}$ erg s$^{-1}$ cm$^{-2}$, if we extrapolate the best-fit CDF log(N)-log(S) with a slope of -1.74. However, if we assume a slope of -1.5, as suggested in our ChaMP sample, galaxies do not dominate until a significantly fainter flux, $f_X$ (0.5-2.0 keV) ~ 2 x $10^{-18}$ erg s$^{-1}$ cm$^{-2}$. We note that a simple extrapolation of the Euclidean slope to faint, high $z$ galaxies may not work. Because the effective volume elements are much smaller at high $z$ than the Euclidean volume elements, the numbers would be smaller and the slope would be flatter, possibly implying that the normal galaxy dominance may occur at an even lower flux. This has particular relevance for the design capacities of future X-ray missions that seek to probe X-ray signatures of star formation at cosmological redshifts.

3.4. X-ray Luminosity Function of Normal Galaxies

With the same sample of galaxies used in determining log(N)-log(S), we determine the X-ray luminosity function (XLF) of normal galaxies. Using the $1/V_a$ method (Schmidt 1968), the source density in a given luminosity bin is given by

$$\Phi(L) = \sum_{i=1}^{N} \frac{1}{V_{a,i}}$$

where $N$ is the number of galaxies in a given luminosity bin and $V_{a,i}$ is the accessible volume for galaxy $i$. Following Hogg (1999), the co-moving volume can be written as

$$V_{a,i} = \frac{c\,\Omega_i(fx)}{H_o} \int_{Z\min}^{Z\max} \frac{(1+z)^2 D_{A,i}^2}{(\Omega_M (1+z)^3 + \Omega_\Lambda)^{1/2}}\, dz$$

where $\Omega_i(f_X)$ is the sky coverage for $f_X$ of galaxy $i$, $D_{A,i}$ is the angular diameter distance at redshift z, and $z_{min}$ and $z_{max}$ are the minimum and maximum redshifts possible for a source to stay in the luminosity bin. We compare our XLF with those determined from the NHS data (Georgantopoulos et al. 2005) and the CDF data (Norman et al. 2004) in Figure 9. Norman et al. (2004) present two XLFs with different redshifts. We take one for z < 0.5 which is more appropriate to compare with our data. We convert the CDF XLF to the common energy band (0.5-8.0 keV) assuming $\Gamma_{ph}$=2 (Kim et al. 1992). Our XLF is consistent with both of them within the statistical error. Using a single power-law, we fit the XLF with the best-fit slope of 1.15 ± 0.1.

4. TYPE 2 AGN AND QSO

We classify a type 2 AGN as an NELG with $f_X/f_O > 0.1$ in section 2. Their X-ray luminosity ranges from $L_X(0.5-8 \text{ keV}) = 10^{42}$ erg s$^{-1}$ to $10^{44}$ erg s$^{-1}$ while their redshift ranges from $z=0.1$ to 1.2. As discussed in section 2, their optical luminosities are in a small dynamical range with ~90% having $M_r = -20 - -22$ mag, which produces the apparent tight correlation between $f_X/f_O$ and $L_X$ (see Figure 1b)

While most of these sources do not appear to be obscured in X-rays, there is a group of 18 X-ray sources with an extremely high hardness ratio, HR > 0.5, corresponding to $N_H > 3 \times 10^{22}$ cm$^{-2}$ for a source spectrum of $\Gamma_{ph} = 1.7$. We note that a few of them were only detected in the H-band. 15 of them are among AGN with $f_X/f_O > 0.1$ and their X-ray and optical properties are listed in Table 4. Two are ALGs and will be discussed in the next section. The remaining one (CXOMP J114150.4+660219) has an intermediate $f_X/f_O$ (see Table 4), but correcting the X-ray flux for the internal absorption will put it in the AGN group.

These obscured NELGs fit the expected characteristics of type 2 AGNs, in that both X-ray and optical emission from broad line regions are consistently absorbed. The remaining type 2 AGNs appear to have normal, unabsorbed X-ray spectra, in contrast to the expectation, but this could be due to the different absorption mechanisms operating by gas and dust, separately (see also Silverman et al. 2005). As shown in Kim et al. (2004) and Silverman et al. (2005), we also found a similar trend where more obscured sources are found among X-ray faint sources.

Apparently, we have no X-ray source which belongs to a type 2 quasar, i.e., $L_X > 10^{44}$ erg s$^{-1}$. However, if we correct the heavily absorbed type 2 AGN for intrinsic absorption, the X-ray luminosity increases by a factor 2 to 30 in $L_X$ (0.5-8 keV) for $N_H = 3 \times 10^{22} - 10^{24}$ cm$^{-2}$. Many of those obscured type 2 AGNs listed in Table 4 are indeed candidates for type 2 quasars.

TABLE 4
Obscured Type 2 AGN

| | net counts S | net counts H | r (mag) | z | log$f_X/f_O$ | log$L_X$ (erg s$^{-1}$) | HR | 1σ limit (lower upper) |
|---|---|---|---|---|---|---|---|---|
| $f_X/f_O > 0.1$ | | | | | | | | |
| CXOMP J033803.0-050046 | 1.39 | 9.53 | 23.46 | 1.192 | 0.09 | 43.31 | 1.00 | (0.53 1.00) |
| CXOMP J033934.2-352349 | 1.69 | 14.85 | 21.52 | 0.533 | -0.65 | 42.48 | 1.00 | (0.62 1.00) |
| CXOMP J045356.8-030227 | 5.22 | 19.31 | 20.05 | 0.423 | -1.00 | 42.48 | 0.61 | (0.38 0.80) |
| CXOMP J054240.9-405627 | 2.51 | 52.51 | 21.06 | 0.639 | -0.18 | 43.32 | 0.93 | (0.85 0.98) |
| CXOMP J134359.2+555259 | -- | 25.01 | 20.67 | 0.593 | -0.66 | 42.93 | 1.00 | (0.90 1.00) |
| CXOMP J134415.5+561214 | 14.21 | 90.50 | 18.63 | 0.132 | -0.74 | 42.15 | 0.76 | (0.59 0.95) |
| CXOMP J153311.2-004524 | 2.89 | 25.75 | 17.82 | 0.151 | -0.89 | 42.46 | 0.86 | (0.66 0.96) |
| CXOMP J154424.2+535546 | 5.86 | 30.81 | 18.70 | 0.439 | -0.43 | 43.63 | 0.70 | (0.55 0.82) |
| CXOMP J205602.1-043645 | 2.33 | 13.57 | 21.74 | 0.466 | -0.39 | 42.51 | 0.77 | (0.51 0.97) |
| CXOMP J205618.7-043430 | 5.60 | 41.25 | 21.59 | 0.527 | 0.01 | 43.10 | 0.78 | (0.66 0.87) |
| CXOMP J205624.8-043534 | 2.46 | 42.32 | 19.48 | 0.260 | -0.87 | 42.35 | 0.92 | (0.82 0.98) |
| CXOMP J205642.0-043301 | 0.99 | 19.69 | 20.91 | 0.467 | -0.61 | 42.63 | 1.00 | (0.82 1.00) |
| CXOMP J214010.5-233905 | -- | 27.17 | 20.66 | 0.453 | -0.61 | 42.70 | 1.00 | (0.88 1.00) |
| CXOMP J214019.1-234838 | 7.14 | 27.29 | 21.59 | 0.387 | -0.13 | 42.64 | 0.61 | (0.41 0.79) |
| CXOMP J234813.3+005611 | 4.86 | 32.76 | 21.50 | 0.550 | -0.45 | 42.72 | 0.77 | (0.62 0.89) |

$f_X/f_O = 0.01 - 0.1$

```
CXOMP J114150.4+660219     --    25.84  19.18  0.235  -1.66  41.58  1.00 (0.88 1.00)
--------------------------------------------------------------------------------
```

## 5. XBONG

Although the existence of an unusual population of X-ray bright, optically normal galaxies (XBONG) has been known since the Einstein mission (Elvis et al. 1981), they are recently attracting more attention on both theoretical and observational grounds as Chandra and XMM-Newton observations reveal a significant number of XBONG candidates (Fiore et al. 2000; Comastri et al. 2002; Georgantopoulos and Georgakakis 2005). It is defined as an X-ray luminous object with no obvious signature of AGN activity, or conventionally as an ALG with $L_X > 10^{42}$ erg s$^{-1}$. Theoretically, it is interesting because XBONG could be an intrinsically luminous, but heavily obscured AGN, where the obscuration must be of large enough covering factor that neither broad nor narrow lines escape.

In our sample, there are 21 XBONG candidates among ALGs with $L_X > 10^{42}$ erg s$^{-1}$ (see Table 1). The X-ray luminosity (corrected for Galactic absorption, but not for intrinsic absorption) of 21 XBONG candidates ranges from $L_X = 10^{42}$ erg s$^{-1}$ to $5 \times 10^{43}$ erg s$^{-1}$, while their redshift ranges $z = 0.1$-$0.8$. In contrast to expectation, most XBONG candidates show no sign of intrinsic absorption, indicating that the absorption model does not explain the majority of these sources. Georgantopoulos and Georgakakis (2005) and Hornschemeier et al. (2005) have reported similar results. However, we found significant excess of X-ray absorption (with HR > 0.6 in 1σ error) in two sources (see Figure 2). Their optical and X-ray properties are listed in Table 5. If they are indeed extremely obscured, their X-ray luminosity will increase by a factor of 10-100 and they could be as luminous as unabsorbed broad-line quasars.

TABLE 5
Obscured XBONG

```
--------------------------------------------------------------------------------
    Name              net counts     r       z    log fx/fo  log Lx     HR    1σ limit
                       S     H     (mag)                    (erg s⁻¹)      (lower upper)
--------------------------------------------------------------------------------
CXOMP J105626.8-033721  14.46 86.04  21.26   0.643   -0.08    43.36   0.72 (0.63 0.81)
CXOMP J205601.2-042955   0.68  9.18  21.33   0.370   -0.74    42.09   1.00 (0.68 1.00)
--------------------------------------------------------------------------------
```

Another possible explanation for XBONG is the dilution of nuclear emission lines by the starlight of the host galaxy (Moran, Filippenko & Chornock 2002), i.e., type 2 AGNs with optical stellar light bright enough to outshine the AGN signature. Seven ALGs (out of 21) with $L_X > 10^{42}$ erg s$^{-1}$, but with $f_X/f_O = 0.01$-$0.1$ could belong to this type. However, for the remaining ALGs with $L_X > 10^{42}$ erg s$^{-1}$, but with $f_X/f_O > 0.1$, stellar dilution is unlikely, because their X-ray, optical luminosities and the ratios are similar to those of type 2 AGNs (see blue and red filled squares in Figure 1). Alternatively, Yuan and Narayan (2005) suggested XBONGs might be explained by an inner radiatively inefficient accretion flow, which could in turn produce relatively strong inverse Compton X-ray emission.

## 6. E+A GALAXIES

Because E+A galaxies have strong Balmer absorption lines with no emission in [OII], they are considered to be post-starburst galaxies which had a starburst less than ~1 Gyr ago, but exhibits no sign of current star formation (e.g., Dressler and Gunn 1983). It is still mysterious what truncated star formation so abruptly and whether they are linked to other types of galaxies in their evolutionary path. Recently, Goto (2005) investigated the morphology of E+A galaxies selected from the SDSS DR2 and suggested that E+A are created by dynamical merger/interaction with a nearby galaxy (see also Blake et al. 2005). Another possible explanation for this phenomenon is a heavily obscured starburst where emission lines associated with the on-going star formation are invisible (e.g., Smail et al. 1999). With the X-ray data, we can search for sub-structures related to the merger/interaction signature and we can also test whether their X-ray spectra are absorbed. In particular, the latter test is critical because the two hypotheses (merger/interaction vs. obscuration) predict very different X-ray spectra: very soft spectra for merging/interaction systems and very hard spectra in obscured systems.

In our sample of ALGs, we measured hydrogen Balmer and [OII] lines to identify E+A galaxies. Following Goto (2005) and Blake et al. (2005), we applied (1) EW (H$\delta$) > 5.5Å (in absorption) and (2) EW ([OII] $\lambda$3727) < 2.5Å (in emission) and EW (H$\alpha$) < 3Å (in emission) for the E+A selection criteria. If H$\alpha$ is not available, we used a combination of other Balmer lines (see Blake et al. 2005).

While there is no E+A galaxy among the normal galaxies with $f_X/f_O$ < 0.01, two E+A galaxies are found with $f_X/f_O$ > 0.01: one among the intermediate $f_X/f_O$ (0.01 – 0.1) group and another among the AGN group with $f_X/f_O$ > 0.1. They are listed in Table 5. Although we have only two E+A candidates, it consists of ~5% of our ALG sample. Both E+A candidates have stronger X-ray emission for their optical flux than those of typical normal galaxies, suggesting that E+A phenomena may enhance the X-ray emission.

Table 6
E+A Galaxies

| name | net counts | | r | z | log fx/fo | log Lx | HR | 1$\sigma$ limit |
| | S | H | (mag) | | | (erg s$^{-1}$) | | (lower upper) |
|---|---|---|---|---|---|---|---|---|
| CXOMP J113956.1+660553 | 17.5 | 0.0 | 18.90 | 0.376 | -1.96 | 42.25 | -1.00 | (-1.00 -0.82) |
| CXOMP J230243.1+083945 | 136.5 | 53.5 | 19.26 | 0.438 | -0.69 | 43.14 | -0.44 | (-0.53 -0.34) |

Since they are unresolved in the X-ray image (at z~0.4), we cannot directly identify any merger/interaction signature in the X-ray image. However, their X-ray spectra are distinct, both being very soft. See also Davis et al. (2003) who found a galaxy with a post starburst optical spectrum whose X-ray emission is luminous ($L_X$ ~ 2 x 10$^{42}$ erg s$^{-1}$) and soft ($\Gamma_{ph}$ ~ 2.9). Strong soft X-ray emission from the galactic hot bubbles/winds is often found in the local merging/interacting galaxies (e.g., the Antennae; Fabbiano et al. 2003). The X-ray emission may be related to luminous X-ray binaries, e.g., ULXs which are also found in the starburst galaxies (Hornschemeier et al. 2005). In particular, CXOMP J113956.1+660553 is detected only in the S band, indicating that the hot ISM dominates its soft X-ray emission. This rules out the dusty absorption hypothesis. Also it has a close companion at the same redshift, a narrow emission line galaxy ~10′′ (or ~50 kpc) away, which is also an X-ray source (CXOMP J113957.4+660547 at $z$=0.378). Therefore, CXOMP J113956.1+660553 fits well into the scenario of merger/interaction with a closely accompanying galaxy at <100 kpc scale (see Goto 2005). The second candidate, CXOMP J230243.1+083945, also has soft X-ray emission. It has no clear companion in either optical or X-ray images. Its high $f_X/f_O$ suggests its X-ray emission may be contaminated by an AGN. Regardless of AGN contamination, the soft X-ray emission again suggests that it is not obscured. Finally, we note that both galaxies are in Chandra field of which the target is a cluster (MS 1137.5+6625 and CL J2302.8+0844), but the clusters are at different distances and not physically associated with the galaxies. Goto (2005) also concluded that E+A galaxies are not associated with clusters.

## 7. SUMMARY

Using ChaMP X-ray and follow-up optical data, supplemented with the SDSS DR3 data, we select 136 extragalactic X-ray sources without broad optical lines (i.e., non-type 1 AGN/quasar). Their redshift ranges between 0.01 and 1.2. Combining our ChaMP sample with previously studied samples (CDF by Hornschemeier et al. 2003 and NHS by Georgantopoulos et al. 2005), we summarize our results:

- To distinguish normal galaxies and AGNs, we utilize several diagnostics: $L_X$, $f_X/f_O$, X-ray spectral hardness, and optical line ratios. We find that conservatively applying $f_X/f_O$ limits provides the most reliable method to separate two populations: normal galaxies for $f_X/f_O$ < 0.01 and AGNs for $f_X/f_O$ > 0.1. We find only one exception, which apparently has a lower $f_X/f_O$ (< 0.01), but is likely an obscured AGN with AGN-like optical line ratios.

- One fifth of 136 X-ray sources have intermediate X-ray-to-optical flux ratios ($f_X/f_O$ = 0.01 – 0.1). They consist of mixed groups of galaxies and AGNs, with different classifications in different diagnostics. They could be high-z starburst galaxies (among NELG), giant elliptical galaxies with a large amount of the hot ISM (among ALG), or obscured low-luminosity AGNs. We note that samples selected using a single $f_X/f_O$ limit are likely contaminated by these objects.

- The X-ray-to-optical flux ratios of normal galaxies are in the range of log $f_X/f_O$ = -3.5 – -2.0. Interestingly, the lower limit (log $f_X/f_O$ = -3.5) is consistent with the lower limit of the $L_X$ (point sources)–$L_O$ relation found in nearby early type (Kim & Fabbiano 2004) and late type galaxies (Colbert et al. 03). It appears that the minimum $f_X/f_O$ remains constant up to z = 0.1, indicating no evolution up to $z=0.1$ (or look-back time of ~1 Gyr).

- Using a subset of our galaxy sample with complete optical imaging/spectroscopic coverage and applying an X-ray completeness correction, we build an log(N)-log(S) relation for ChaMP normal galaxies. The best-fit slope is 1.5 ± 0.15, which is considerably steeper than that of the general cosmic background source population. This is consistent with previous results, but slightly flatter than that of the CDF sample (Hornschemeier et al. 2003). As Bauer et al. (2004) pointed out, normal galaxies will outnumber AGNs at faint fluxes, but our estimated flux where two populations are equal ($f_X < 2 \times 10^{-18}$ erg s$^{-1}$ cm$^{-2}$) is a factor of ~5 lower than the previous estimate.

- Sixteen NELGs (mostly with $f_X/f_O > 0.1$) are heavily obscured in X-rays and after correcting their intrinsic absorption, approximately half are type 2 QSO candidates with $L_X > 10^{44}$ erg s$^{-1}$.

- Twenty-one ALGs are X-ray luminous with $L_X > 10^{42}$ erg s$^{-1}$, but optically normal galaxies (XBONG). As opposed to the expected obscuration, we find only two significantly obscured objects with an absorption-corrected $L_X$ comparable to broad line quasars. It appears that dilution by the host galaxy may not work for the majority. The remaining XBONG candidates require alternative explanations, e.g., inner radiatively inefficient accretion flow (Yuan and Narayan 2005).

- We have identified two E+A galaxies with strong Balmer absorption lines, but no [OII] line. Their X-ray spectra are soft and one of them has a companion galaxy ~50 kpc away. Our results support the merger/interaction hypothesis for the origin of this unusual phenomenon (Goto 2005), and rule out the dusty absorption hypothesis.

This work was supported by CXC archival research grant AR4-5017X. We acknowledge support through NASA contract NAS8-39073 (CXC).

# Figure Captions

Figure 1. (a) The optical r magnitude is plotted against the X-ray flux (0.5-8keV). (b) The X-ray-to-optical flux ratio ($f_X/f_O$) is plotted against $L_X$. NELGs and ALGs are distinguished by blue and red colors, respectively. Normal galaxies (with $f_X/f_O < 0.01$) and AGNs (with $f_X/f_O > 0.1$) are marked by filled circles and filled squares, respectively. Open circles mark those unclassified objects with intermediate $f_X/f_O$. Open circle with a dot in the center indicate those galaxies with less accurate optical photometric data (see the text). Also plotted are normal galaxies identified in CDF-N (Hornschemeier et al. 2003) and XMM-Newton NHS (Georgantopoulos et al. 2005), marked by cyan and green triangles, respectively. The black dashed line and the red dotted line indicate $f_X/f_O = 0.1$ and 0.01, respectively. The diagonal solid line in Figure (b) indicates a linear relation with slope=1.

Figure 2. X-ray spectral hardness plotted against $f_X/f_O$ (a) without error bars for visibility and (b) with error bars. The hardness ratio and the corresponding error are determined by a new Bayesian technique (see the text). All the symbols are the same as in Figure 1.

Figure 3. Optical line diagnostics. All the symbols are the same as in Figure 1. The curves separating AGNs and galaxies are taken from Kewley et al. (2001).

Figure 4. (a-d) Distributions of X-ray/optical luminosity/flux of our ChaMP sample. Redshift distributions of (e) our ChaMP sample and (f) CDF-N and NHS galaxy samples. While the total sample is marked by green solid lines, NELGs and ALGs are distinguished by blue solid and dashed lines, respectively. Normal galaxies (with $f_X/f_O < 0.01$) and AGNs (with $f_X/f_O > 0.1$) are marked by red solid and dashed lines, respectively.

Figure 5. (a-b) Same as Figure 1a-b, but only for normal galaxies (with $f_X/f_O < 0.01$). The blue and red long-dashed lines in (a) indicate the minimum $f_X/f_O$ determined from local early type galaxies by Kim and Fabbiano (2004) and spiral galaxies by Colbert et al. (2003).

Figure 6. $L_X$-$M_r$ for (a) elliptical galaxies (or ALG) and (b) spiral galaxies (NELG). The dashed lines are the best-fit relations.

Figure 7. (a) $f_X/f_O$ and (b) $L_X/L_B$ as a function of z. The symbols are the same as in Figure 1.

Figure 8. log(N)-log(S) relations of ChaMP normal galaxies determined in the S (red) and B (blue) energy bands. Also plotted for comparison are those determined with the CDF-N (cyan; Hornschemeier et al. 2003) and NHS samples (green; Georgantopoulos et al. 2005).

Figure 9. X-ray luminosity function for ChaMP normal galaxies (red diamonds). Also plotted are XLFs by Georgantopoulos et al. (2005; green circles) and by Norman et al. (2004; cyan squares).

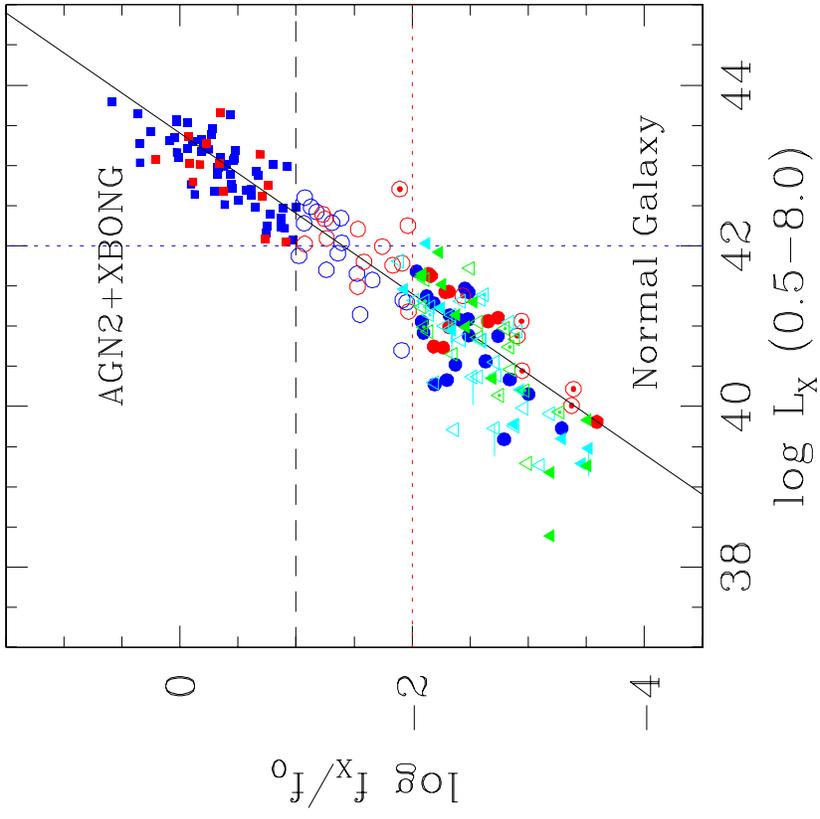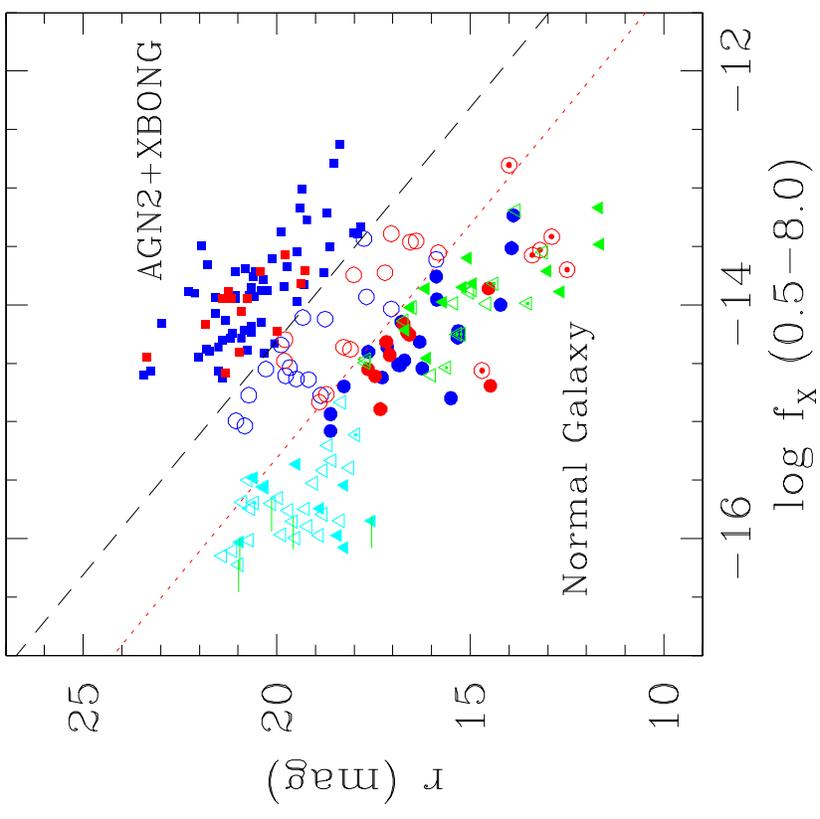

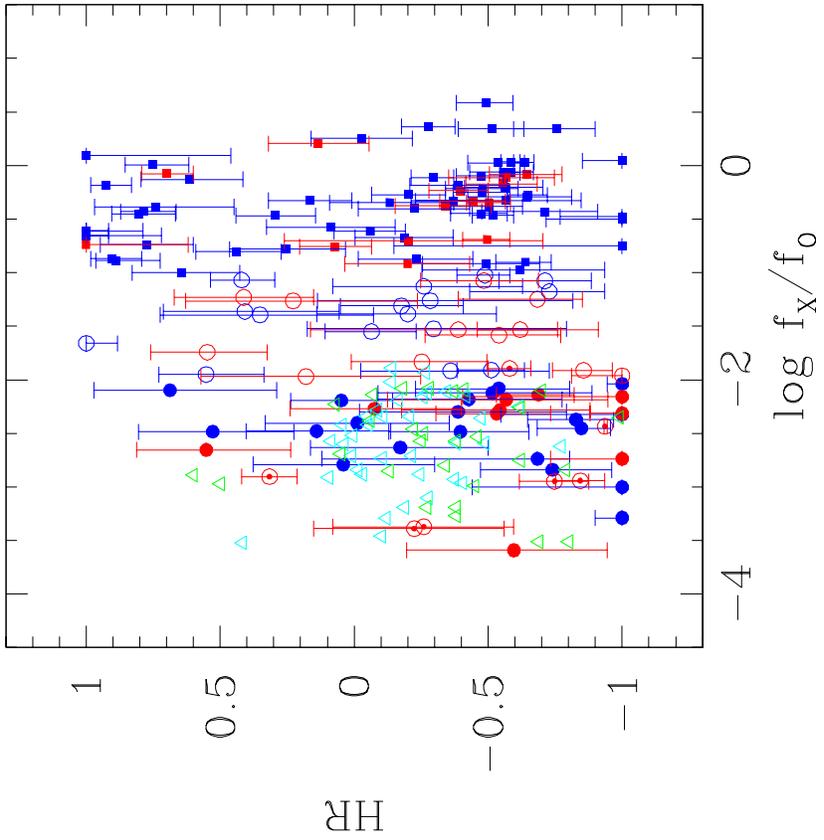
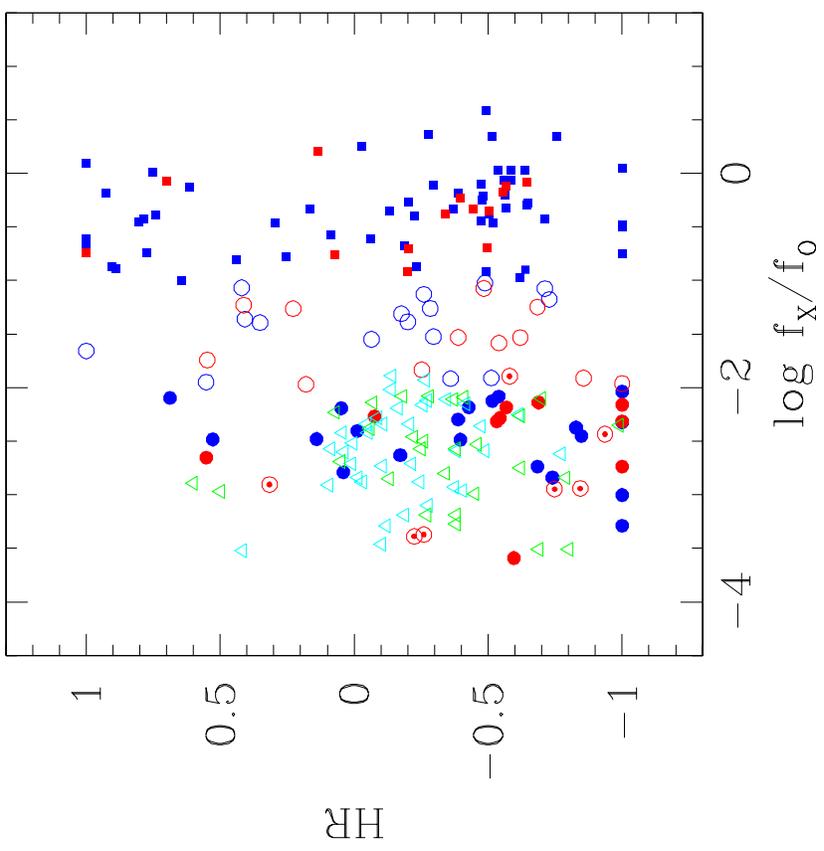

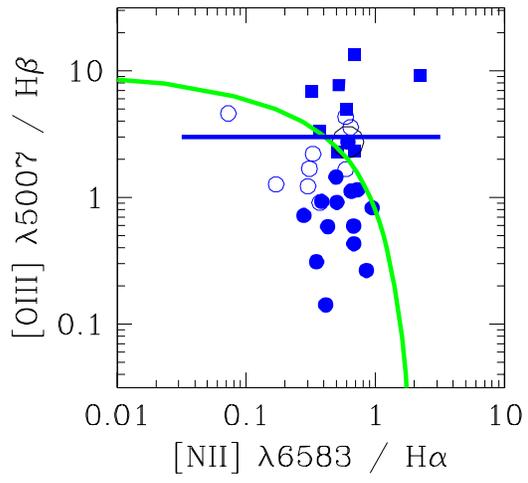

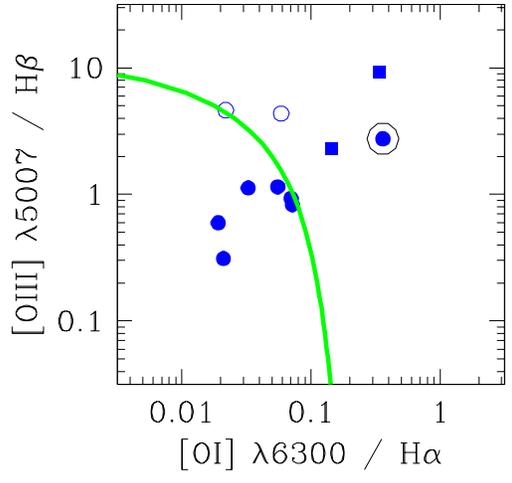

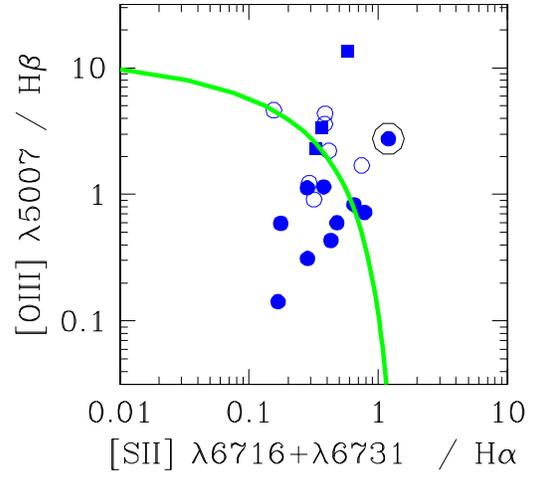

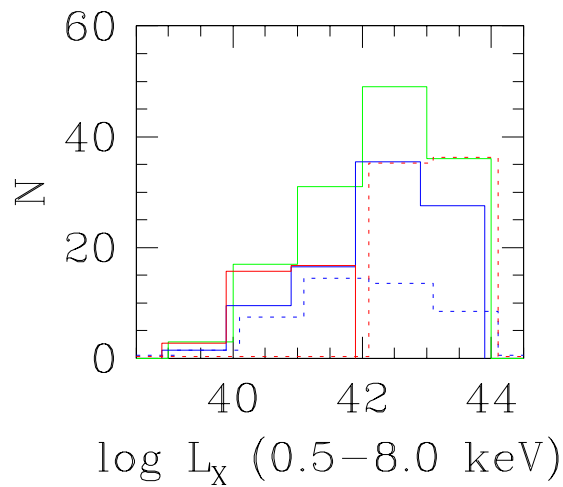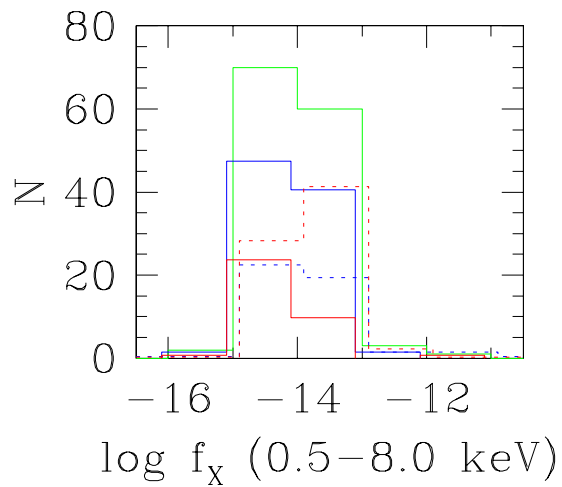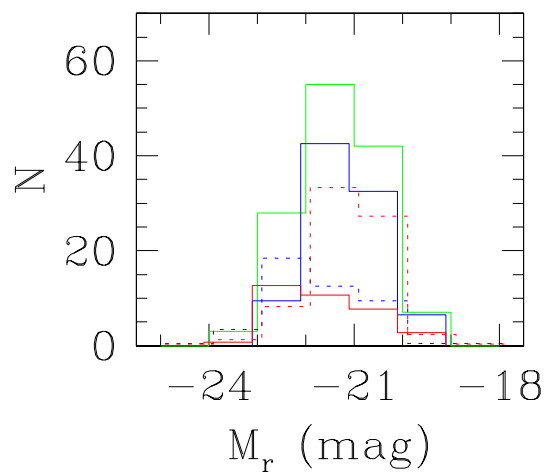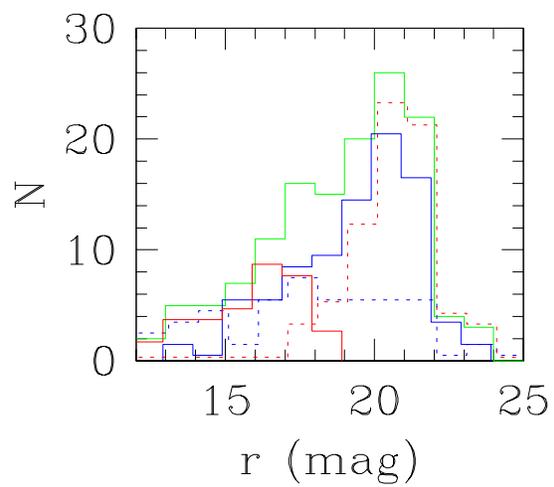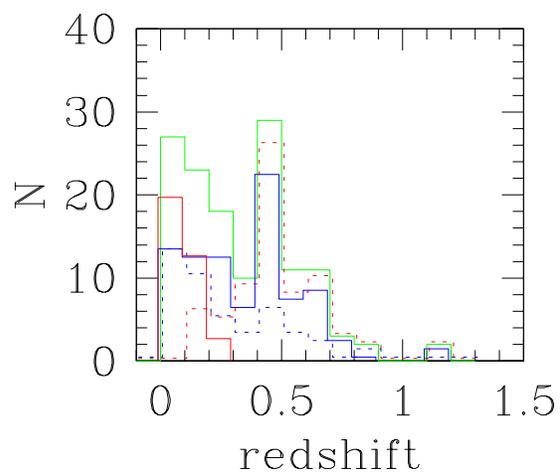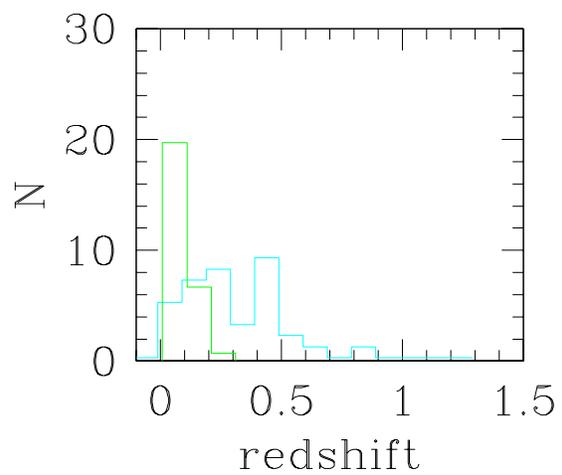

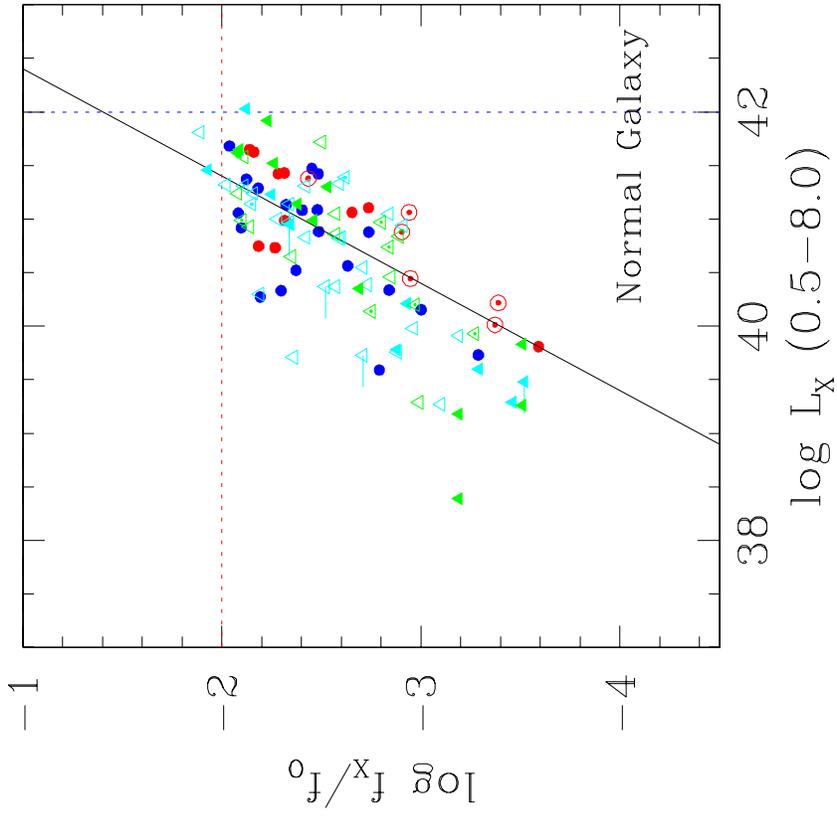
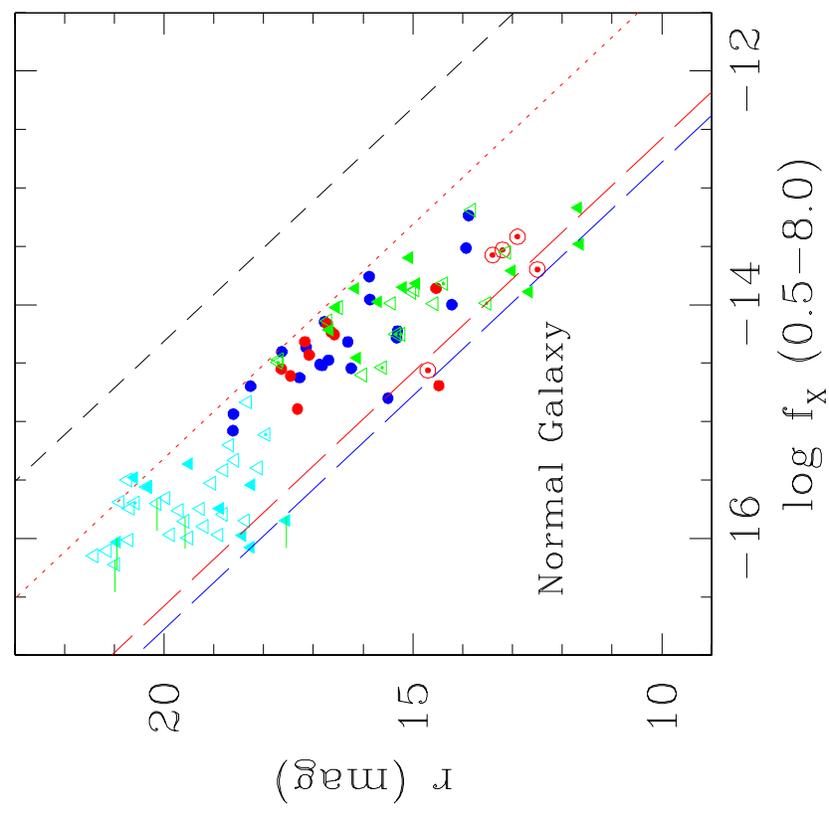

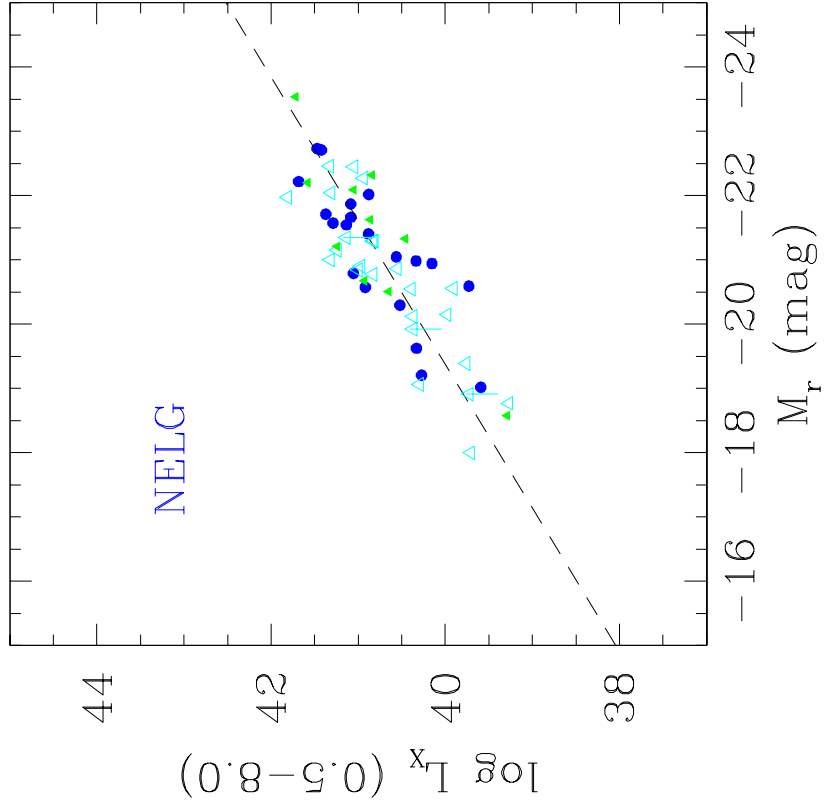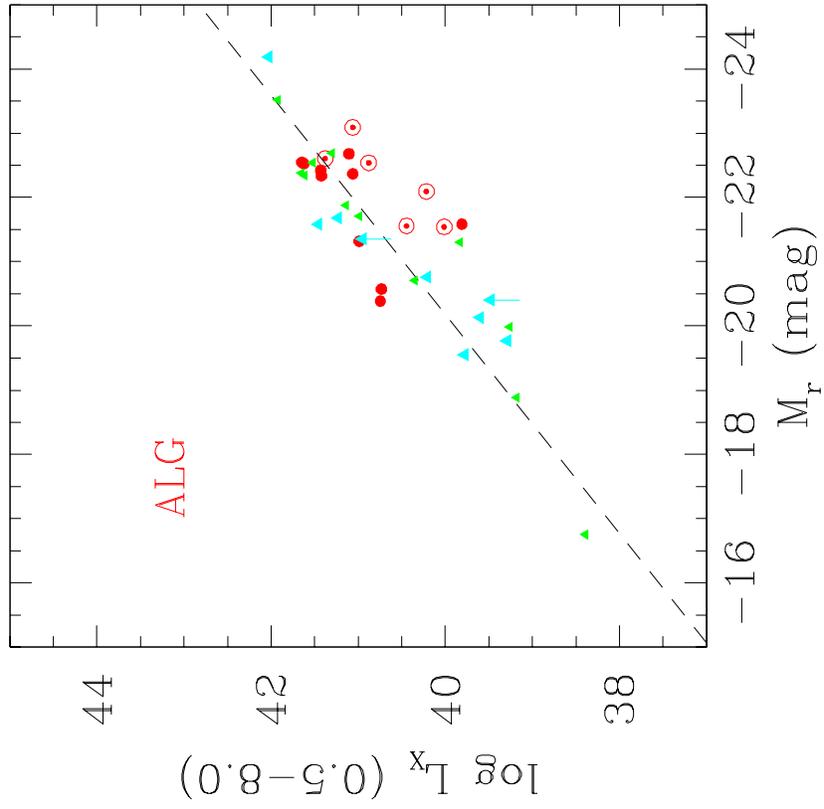

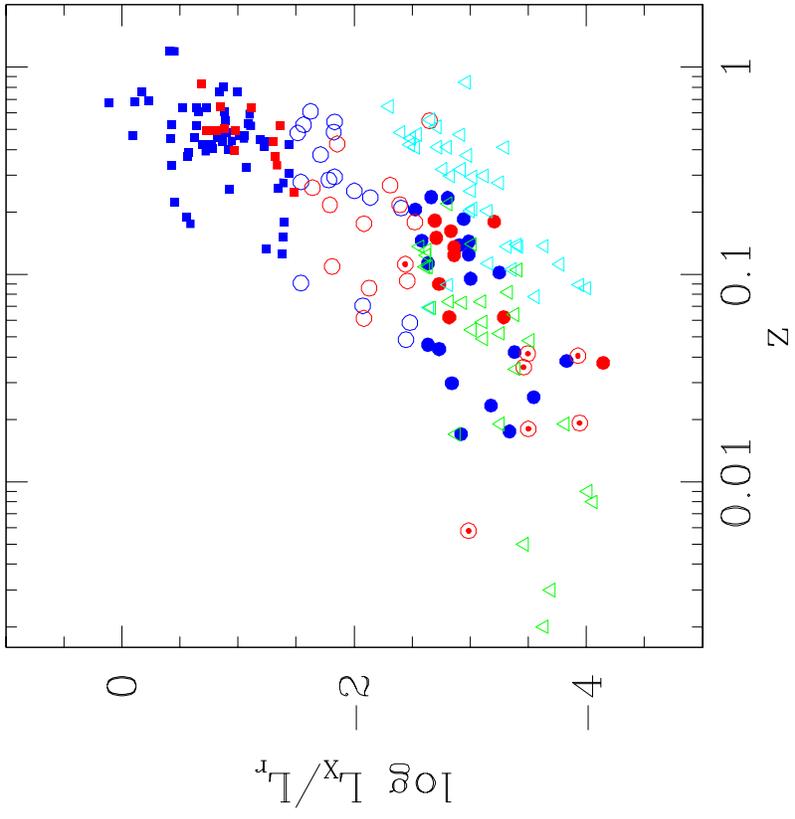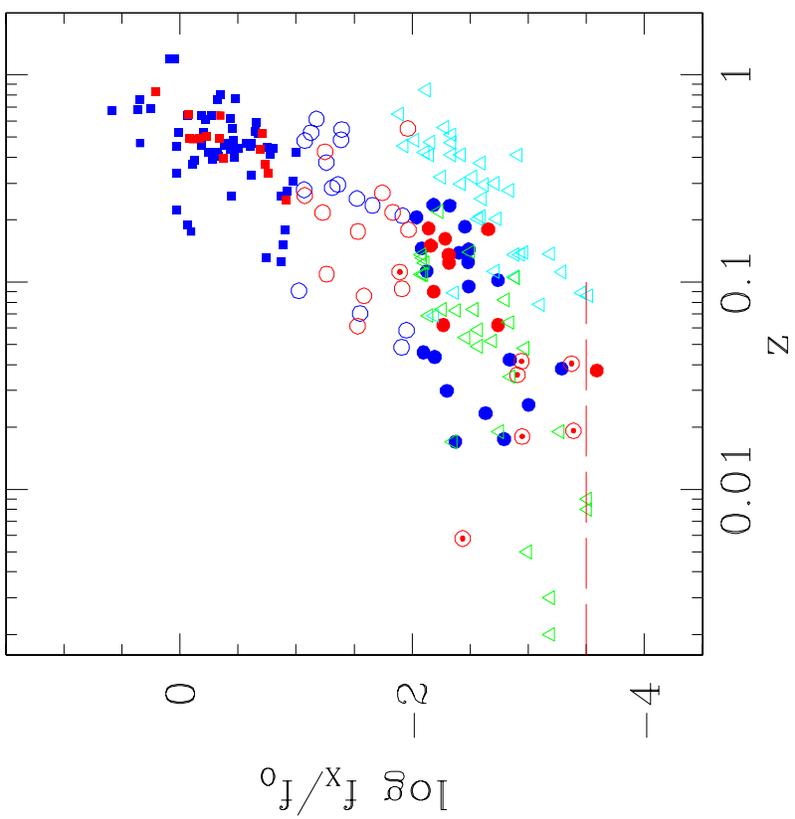

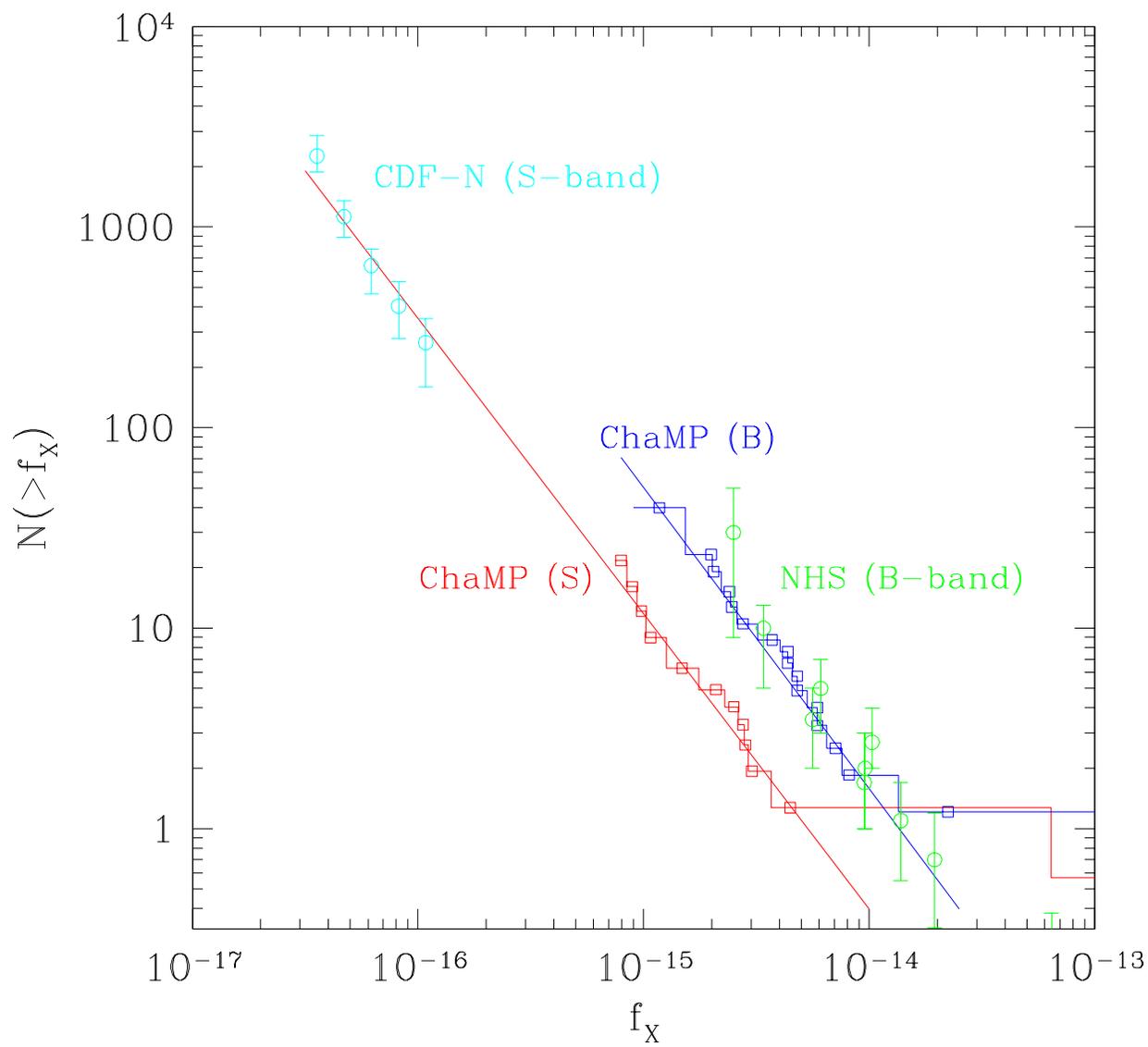

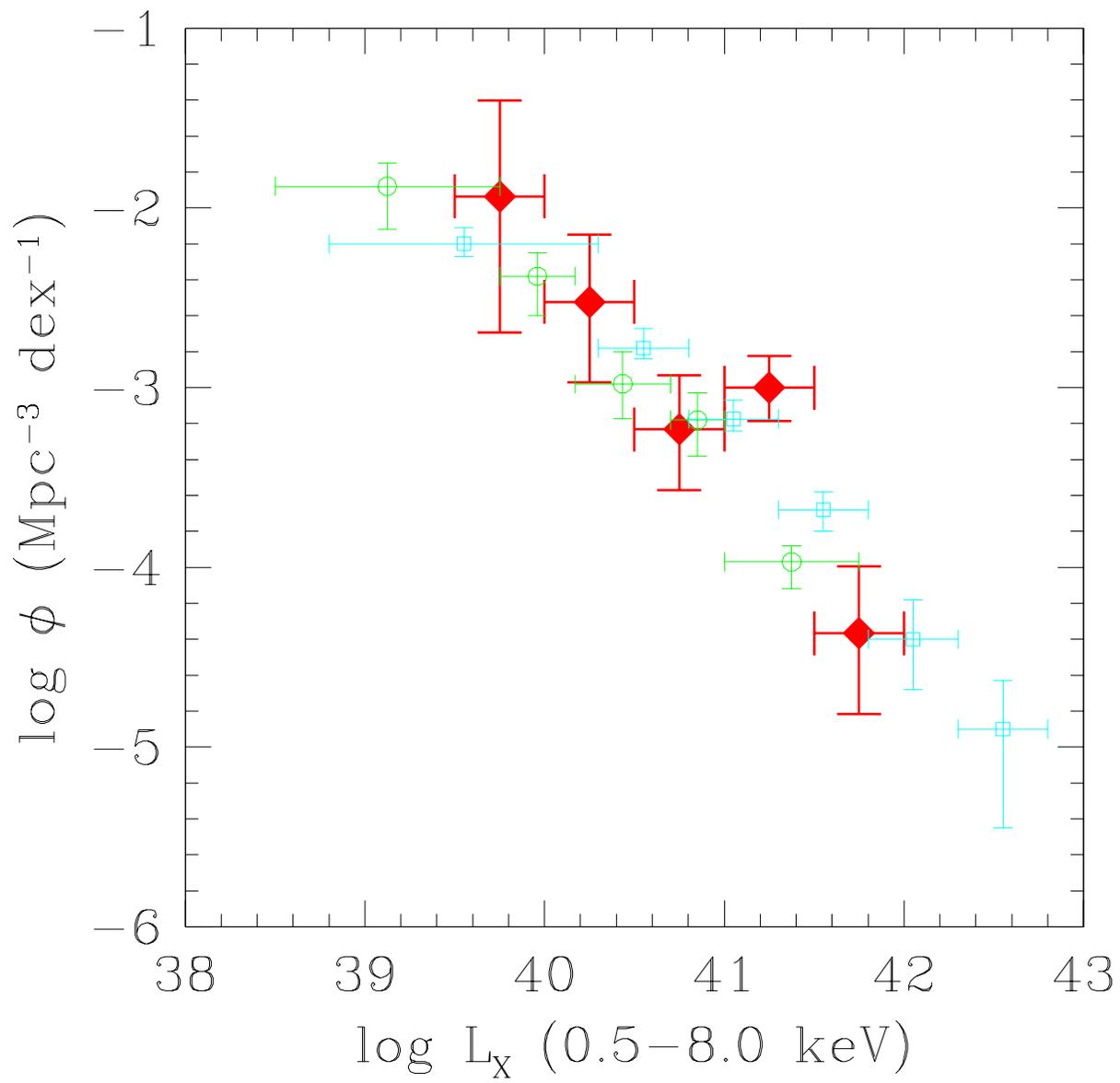